# Strongly first order melting of a two dimensional molecular solid

Rakesh S. Singh, Mantu Santra and Biman Bagchi*

Solid State & Structural Chemistry Unit

Indian Institute of Science, Bangalore 560012, India.

*Email: bbagchi@sscu.iisc.ernet.in

## Abstract

**Melting and freezing transitions in two dimensional systems are known to show highly unusual characteristics. Most of the earlier studies considered atomic systems; the melting behavior in two dimensional molecular solids is still largely unexplored. In order to understand the role of multiple energy and length scales present in molecular systems on nature of melting transition, here we report computer simulation studies of melting of a two dimensional Mercedes-Benz (MB) system. We find that the interplay between the strength of isotropic and anisotropic interactions can give rise to rich phase diagram. The computed solid-liquid phase diagram consists of isotropic liquid and two crystalline phases – honeycomb and oblique. In contradiction to the celebrated KTHNY theory, we observe strongly one step first order melting transitions for both the honeycomb and oblique solids. The defects in both solids and liquids near the transition are more complex compared to the atomic systems.**

For two dimensional (2D) solids, thermal fluctuations spontaneously destroy occurrence of any long range translational order [1]. As a result, two dimensional melting exhibits many unusual features not encountered in three dimensional systems. In fact, even the exact nature of melting of 2D solids is still a matter of debate. Most of the previous studies have focused on atomic systems. The case of melting of solids where molecules interact via both isotropic as well as anisotropic interactions has remained largely unexplored.

The beautiful defect-unbinding (Kosterlitz-Thouless-Nelson-Halperin-Young (KTHNY)) theory for two dimensional melting has generated enormous interest in the last three decades [2-7]. This theory predicts a two step continuous melting transition where in 2D triangular crystal, the unbinding of dislocation pairs into isolated dislocations creates an intermediate hexatic phase, which has quasi-long range six-fold bond orientational order (characterized by algebraic decay of the spatial correlation of six-fold bond orientational order) but no long-range translational order. This transition is followed by a further unbinding of isolated dislocations leading to the true liquid phase having only short range translational as well as bond orientational order.

In some respects, the hexatic phase resembles the nematic liquid crystal, though the later has only two-fold symmetry. Recently, many computer simulation [8-13] and experimental [14-18] studies have confirmed the existence of anisotropic intermediate phase, although many questions still remain unanswered. Does the melting of *molecular solids* in 2D also show the same dislocation-unbinding melting behavior? If yes, what are the nature of transition and the order of intermediate phase? Is there any effect of presence of multiple length and energy scales in molecular systems on the nature of melting transition? Does the extent of specificity in the interaction have any effect on the nature of melting transition? In order to provide the answer to few of these questions, in this work, we have studied the melting transition of a model molecular system interacting via both isotropic as well as anisotropic interactions. This model is popularly known as Mercedes-Benz (MB) model and show many remarkable water-like anomalies for specific value of parameters [19-21] (details of the potential are discussed in the methods section).

The systems of particles interacting via two length scale isotropic potential are well studied and known to show many unusual phenomena (such as polymorphism in both crystal and glassy phases, liquid-liquid transition and other water-like anomalies, anomalous melting at high pressure) not encountered in systems having one length scale potential [22-27]. In general, many unusual phenomena in complex systems can be attributed to the interplay between multiple length and energy scales. The case of molecular systems, where the anisotropic interactions play important role is much more complex. In molecular systems, particles interact via both short range isotropic as well as long range anisotropic interactions having multiple energy scales. Recently, in SPC/E water model, by modifying the strength of Lennard-Jones (LJ) term relative to electrostatic term Debenedetti *et al*. [28] found that the competition between spherically and tetrahedral symmetries can lead to an intermediate regime where the fluid is less structured than at either the tetrahedral or spherically symmetric extremes. In the next section we discuss the phase diagram of a two dimensional molecular MB system having two length as well as energy scales with varying strength of isotropic to anisotropic interaction.

In **Figure 1(A)** we show the computed phase diagram for the melting transition of MB system. Here we have plotted the transition points for different relative strengths of isotropic to anisotropic interaction ($\lambda_{an} = \varepsilon_{LJ} / \varepsilon_{HB}$) at two different pressures P = 0.10 (red line with triangle up symbols) and P = 0.19 (black line with circles).

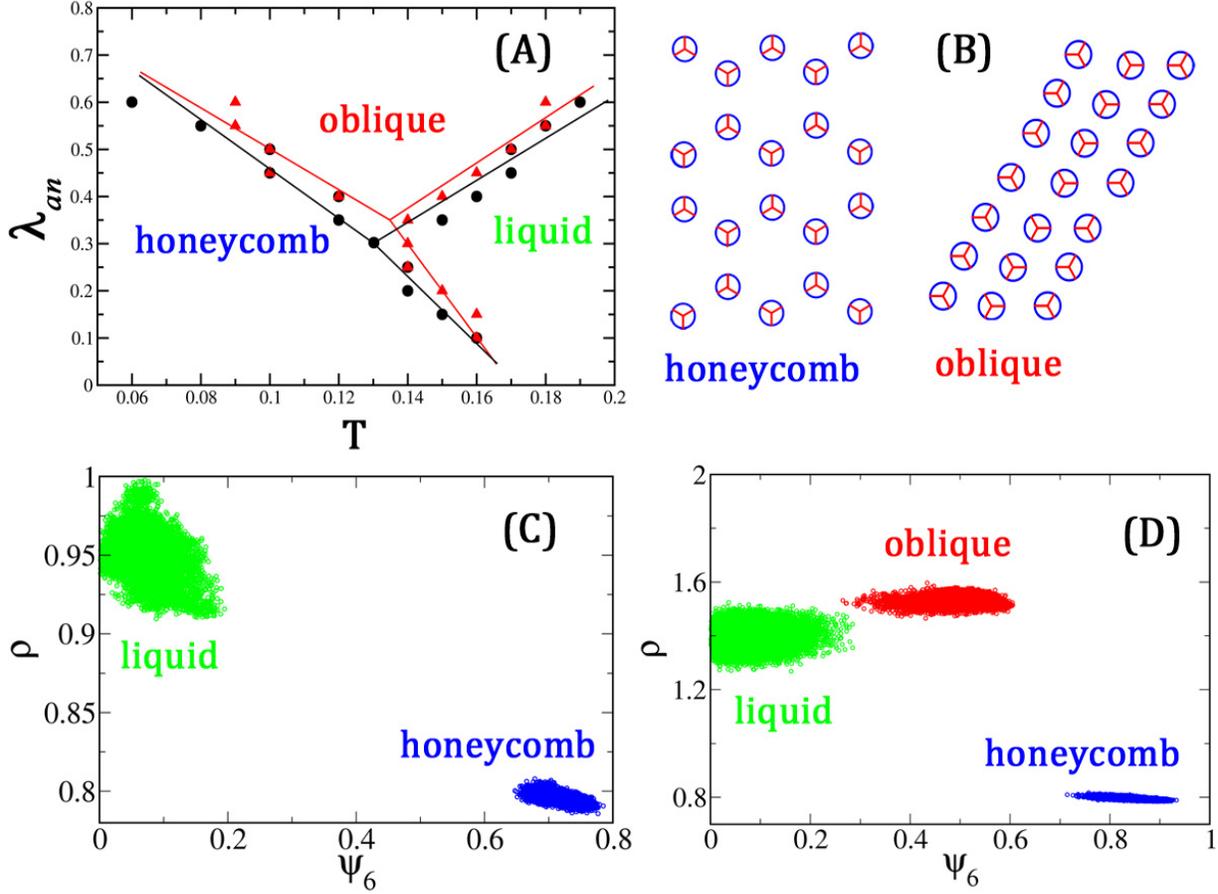

**Figure 1. (A)** Solid-liquid and solid-solid transition lines obtained for the two dimensional Mercedes Benz (MB) model potential. Loci of transition points for liquid-solid and solid-solid phase transition(s) for different strengths of isotropic to anisotropic interaction ($\lambda_{an}$) at two different pressures, P = 0.10 (red line with triangle up symbols) and P = 0.19 (black line with circles). **(B)** The structures of the honeycomb and oblique solid phases are shown. The alignment of bonds shows hydrogen bonding. **(C)** A two dimensional projection of the configurations along density (ρ) and six-fold bond orientational order parameter ($\psi_6$) for $\lambda_{an} = 0.1$ *(weak isotropy)*, and **(D)** the same as in (C) but for $\lambda_{an} = 0.4$ *(strong isotropy)* at P = 0.19. In the system, particles interact via both an isotropic Lennard-Jones (LJ) and an explicit anisotropic hydrogen bond (HB) term such that the interaction energy is written as $U(\mathbf{X_i}, \mathbf{X_j}) = U_{LJ}(r_{ij}) + U_{HB}(\mathbf{X_i}, \mathbf{X_j})$, $\mathbf{X_i}$ denotes vectors representing both the coordinates and orientation of $i^{th}$ particle and $r_{ij}$ is the distance between the centers of two particles. Note that we have two length as well as two energy scales corresponding to **the isotropic LJ and anisotropic HB interactions. The explicit form of potential is discussed in methods section.**

At low values of $\lambda_{an}$ (high anisotropy) a sharp one step transition leads the low density honeycomb solid phase to high density liquid phase. *This transition bears close resemblance to the melting of ice.* On increasing $\lambda_{an}$ (that is, on increasing the strength of isotropic interaction), we find a bifurcation of this single transition line into two transition lines where isotropic high density liquid and the low density honeycomb solid phase (HSP) are separated by an oblique solid phase (OSP). The bifurcation of the transition line arises due to the interplay between the ground states of isotropic (triangular solid) and anisotropic (honeycomb solid) parts of the interaction potential. *At the triple point both the isotropic and anisotropic interactions balance each.* Note that this triple point is different than the true triple point where three coexistence lines intersect. Above the triple point, isotropic interaction dominates over the anisotropic part and thus promoting a switch in the ground state of system (from honeycomb to oblique) occurs. It is interesting to note that the oblique phase is neither the ground state of isotropic part nor the anisotropic part of the potential. Oblique phase is stabilized due to interplay between isotropic and anisotropic part of interaction. The detailed structures of honeycomb and oblique solids are provided in **Fig. 1(B)**. The honeycomb structure is stabilized by three anisotropic hydrogen bond (HB) interactions per particle; however, in oblique phase particles are stabilized by (two) HBs as well as Lennard-Jones (LJ) interactions. The alignment of bonds in the figure shows the hydrogen bonding. We note that the emergence of two length scales and decrease in extent of specificity of interaction in the ground state of the potential in going from low isotropic to large isotropic interaction side in the phase diagram.

On increasing pressure, the transition lines between the low density honeycomb and the high density phases (both the isotropic liquid or crystalline oblique) shift towards lower temperature, as the high density phase gains stability on increasing pressure. On the other hand, on increasing pressure, the transition line between oblique solid to the liquid shifts towards the higher temperature. *Thus on increasing pressure, the range of temperature of stability of the oblique lattice increases.* This shift is easily understood by using the well-known Clausius-Clapeyron equation, $dP/dT = \Delta S / \Delta V$, where $\Delta S$ is the change in the entropy and $\Delta V$ is the change in volume from the initial phase to the final phase.

In **Fig. 1(C)** and **1(D)**, we show the range of stability of the phases in a two dimensional order parameter (the six-fold bond orientational order and the density) space. Note the larger fluctuations allowed along the bond orientational order for the honeycomb and oblique solids compared to that along density. In the subsequent sections we shall discuss the melting transition for two cases (i) honeycomb to liquid ($\lambda_{an} = 0.1$), where transition is accompanied by increase in density and closely resembles with melting of ice and (ii) oblique to liquid ($\lambda_{an} = 0.4$), where liquid phase has lower density than the corresponding solid phase. These two solid phases (honeycomb and oblique) differ in the extent of specificity in the inter-particle interaction.

The most fundamental step towards understanding the nature of transition is to study the underlying free energy surface and its dependence on temperature. **Figure 2** shows the computed one dimensional (at different temperatures) and two dimensional free energy surfaces (near transition point) for both $\lambda_{an} = 0.1$ (honeycomb to liquid transition) and $\lambda_{an} = 0.4$ (oblique to liquid transition). The free energy barrier near coexistence for $\lambda_{an} = 0.1$ (**Fig. 2A**) is comparatively larger than $\lambda_{an} = 0.4$ (**Fig. 2C**). This indicates that surface energy of honeycomb-liquid interface is larger than the oblique-liquid interface. We also note that the curvatures of solid and liquid basins in the free energy surface are significantly different for $\lambda_{an} = 0.1$. Large curvature of the solid basin signifies the absence of large scale density fluctuations, whereas for liquid basin the opposite is true. In the case of $\lambda_{an} = 0.4$ (oblique to liquid transition, **Fig. 2(C)**), the basin corresponding to the solid phase in the free energy surface is comparatively flatter than the case of $\lambda_{an} = 0.1$.

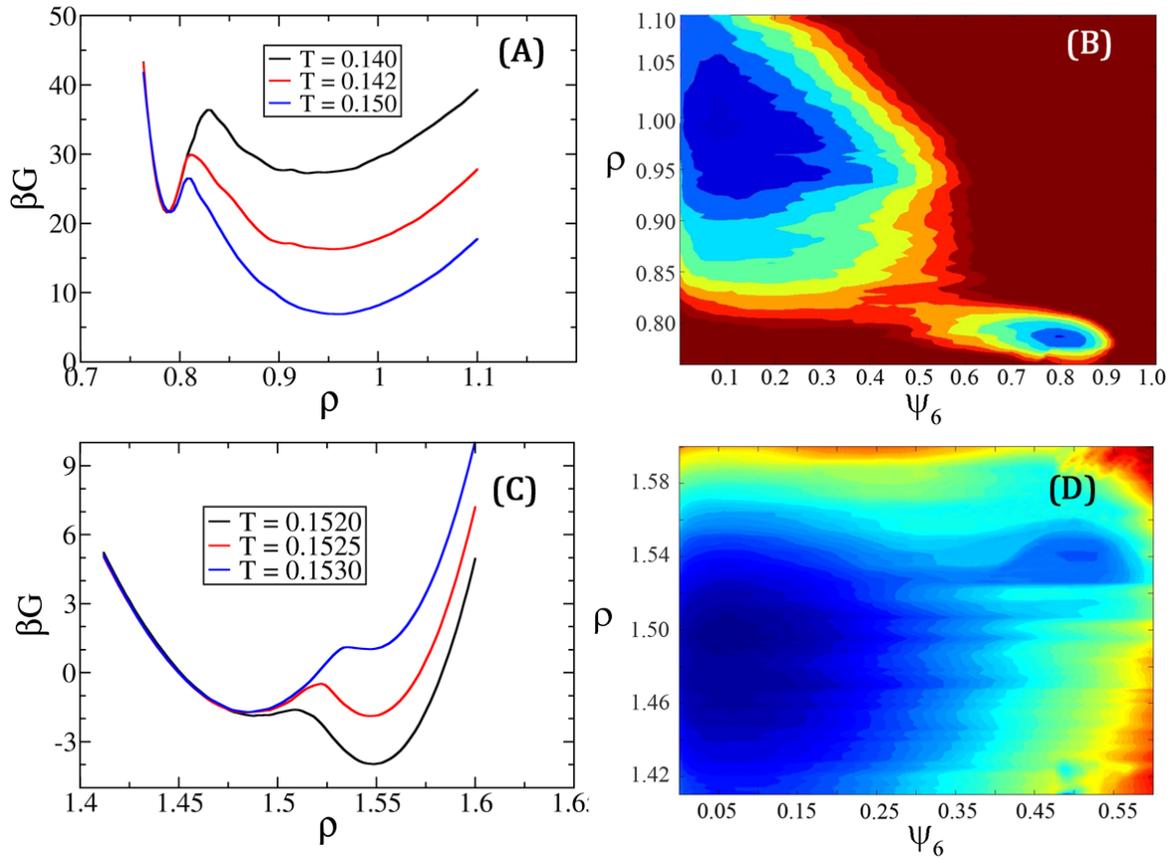

**Figure 2. The temperature evolution of free energy surface of melting is shown. (A) Gibbs free energy surface ($\beta G$) for melting transition is plotted against density (ρ) as an order parameter for (A) $\lambda_{an} = 0.1$ (honeycomb to liquid) and (C) $\lambda_{an} = 0.4$ (oblique to liquid) at three different**

**temperatures. A projection of two dimensional free energy surface for melting with six-fold bond orientational order (ψ₆) and density (ρ) as order parameters is shown for (B) $\lambda_{an} = 0.1$ and (D) $\lambda_{an} = 0.4$ near respective transition temperatures. The free energy is computed for a system having 240 molecules using Transition Matrix Monte Carlo (TMMC) [31]. The curvature of honeycomb solid basin is significantly different from the corresponding liquid basin as well as oblique solid basin. The flatness of the free energy basin for liquid indicates the presence of large scale density and order fluctuations.**

The flatness of the free energy surface can be correlated to the relative strength of anisotropic interaction or the extent of the specificity in the inter-particles interactions (honeycomb lattice has three specific interactions per particle; however, oblique has only two). Thus there is a direct relation between the stiffness of the free energy surface (and height of free energy barrier) and the relative strength of anisotropic interaction. This has important consequences on the nature of melting transition that we shall discuss later. In order to show the pathway of transition in order parameter space, in **Fig. 2(B)** and **2(D)**, we have shown the two dimensional free energy surface considering density ($\rho$) and six-fold bond orientational order ($\psi_6$) as two relevant order parameters near respective transition temperatures.

In **Fig. 3(A)** and **3(B)** we report the density and order (both translational and bond orientational) change on melting for $\lambda_{an} = 0.1$ and in **Fig. 3(C)** and **3(D)** we report the same for $\lambda_{an} = 0.4$. Contrary to the atomic two dimensional systems where the total density change (decrease) on melting/freezing is < 5%, in this model for $\lambda_{an} = 0.1$, we observe a change (increase) of ~ 25%. This change arises due to the collapse of the low density honeycomb structure to the high density liquid and as mentioned earlier show close resemblance with the melting of ice. For the case of $\lambda_{an} = 0.4$, the density change (~5%) is almost same as in the case of melting of atomic systems.

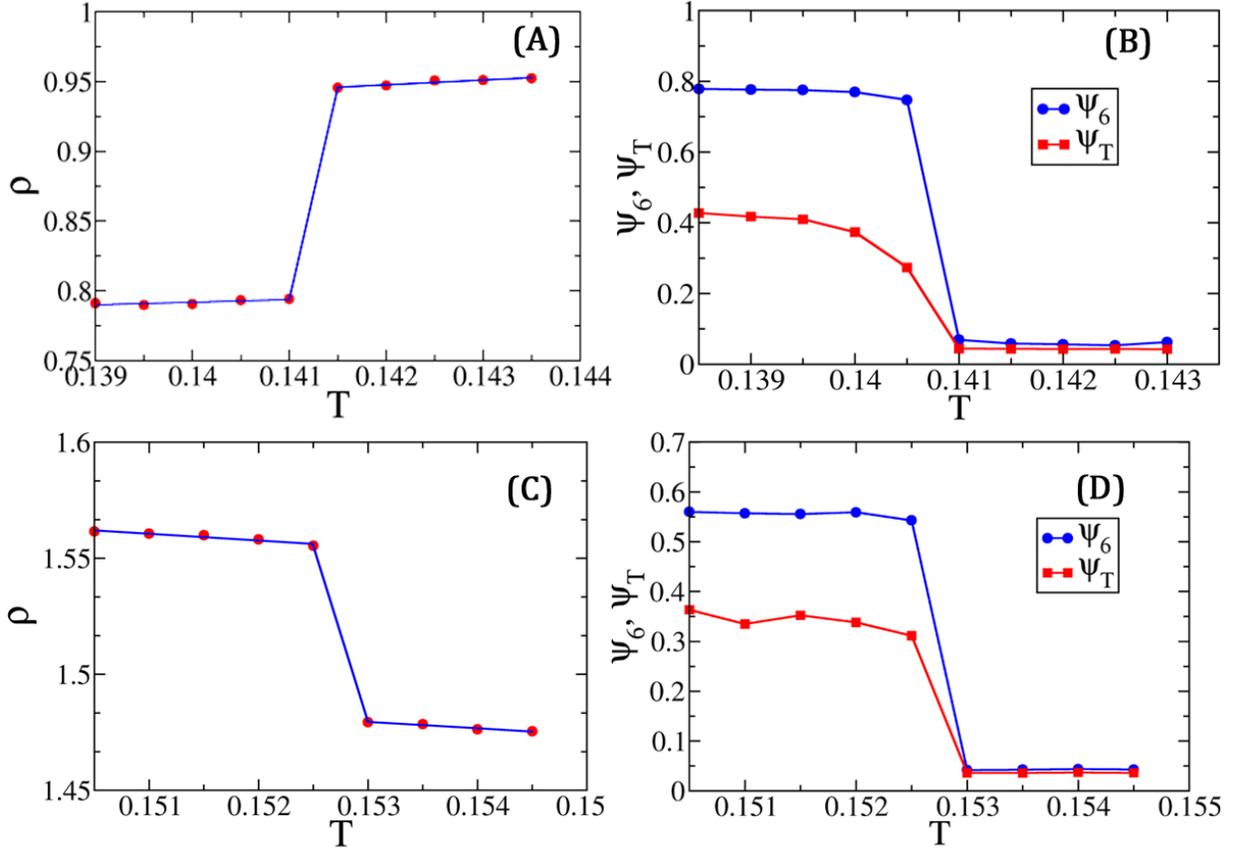

**Figure 3.** Temperature dependence of density ($\rho$), translational ($\psi_T$) and bond orientational order parameter ($\psi_6$) are shown. (A) Density change for $\lambda_{an} = 0.1$, (B) $\psi_6$ and $\psi_T$ change for $\lambda_{an} = 0.1$. (C) Density change for $\lambda_{an} = 0.4$, (D) $\psi_6$ and $\psi_T$ change for $\lambda_{an} = 0.4$. For $\lambda_{an} = 0.1$, the starting configuration is honeycomb solid and oblique solid for $\lambda_{an} = 0.4$ at P = 0.19.

**Figures 3(B)** and **3(D)** show the change in orientational and translational orders with temperature for $\lambda_{an} = 0.1$ and 0.4. For the oblique crystal to liquid transition ($\lambda_{an} = 0.4$) we have defined $\psi_6$ as an order parameter. For the case of $\lambda_{an} = 0.1$, since the honeycomb lattice has 3-fold symmetry and global $\psi_3$ becomes zero (due to cancelation of components for odd symmetry), global $\psi_3$ does not distinguish liquid and solid phases. One can use a local order parameter, defined as $\psi_{3l} = \sum_{i=1}^{N} |\psi_{3i}|/N$. However, it is also not a good choice as the liquid phase has also significant local solid-like ordering. We have thus used $\psi_6$ as an order parameter to distinguish the honeycomb solid and the liquid state. As depicted in **Fig. 3(B)** and **3(D)** we observe a sharp jump at same temperature for both the global bond orientational order ($\psi_6$) as well as translational order parameter ($\psi_T$).

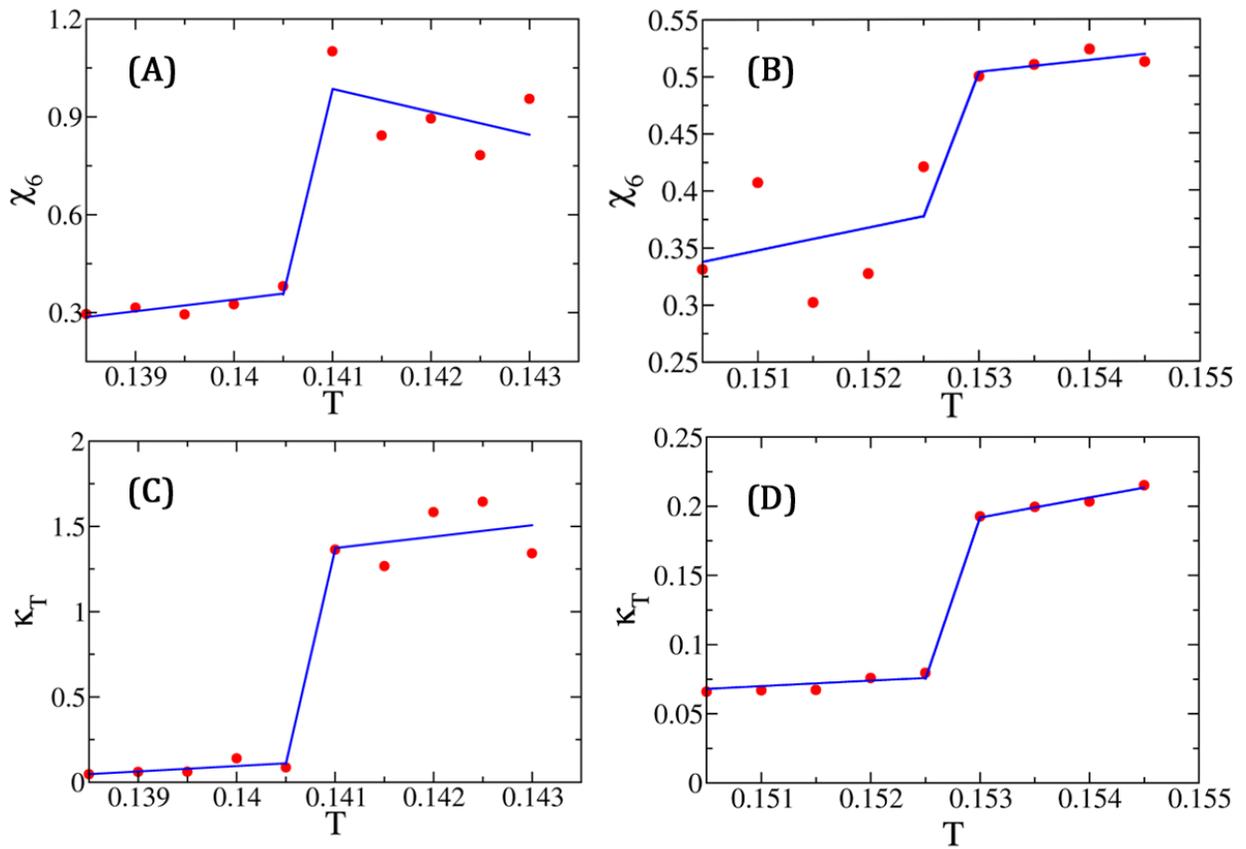

**Figure – 4.** Temperature dependence of the bond orientational order susceptibility ($\chi_6$) for **(A)** $\lambda_{an} = 0.1$, **(B)** $\lambda_{an} = 0.4$ and isothermal compressibility $(\kappa_T)$ for **(C)** $\lambda_{an} = 0.1$ , **(D)** $\lambda_{an} = 0.4$ is shown. Both the bond orientational order susceptibility ($\chi_6$) and isothermal compressibility

**$\left(\kappa_T\right)$ show weak temperature dependence and finite jump at transition temperature. This indicates one step 1<sup>st</sup> order melting transition and contradicts the KTHNY predictions.**

The temperature dependence of susceptibilities of order parameters provides more insight into the nature of transition. As depicted in **Fig. 4,** both the bond orientational order susceptibility ($\chi_6$) and compressibility ($\kappa_T$) show weak temperature dependence and finite jump indicating a strongly 1<sup>st</sup> order phase transition and contradicts the KTHNY predictions.

A more direct evidence for the presence of any intermediate phase during melting can be gained by computing both, the bond orientational order correlation function, $g_6(r)$ and translation order correlation function, $g_G(r)$ (which is a measure of extent of spatial correlation of six-fold bond orientational order and correlation in translation of lattice points). In **Fig. 5(A)-(D)** we have plotted the computed $g_6(r)$ and $g_G(r)$ at different temperatures (near transition point) for both $\lambda_{an} = 0.1$ and $\lambda_{an} = 0.4$. For $\lambda_{an} = 0.1$, we observe that below melting temperature there is a long range bond orientational order and quasi-long range translation order (characterized by algebraic decay of $g_G(r)$). On increasing temperature, we observe a transition from long range bond orientational order and quasi-long range translational order to short range correlation of both bond orientational order as well as translational order (exponential decay of $g_6(r)$ and $g_G(r)$). Absence of any intermediate quasi-long-range orientational correlation (quantified by the algebraic decay of $g_6(r)$) implies the melting as a one step process. However, for $\lambda_{an} = 0.4$, near transition temperature, we observe a weakly long range bond orientational order and quasi-long range translational order.

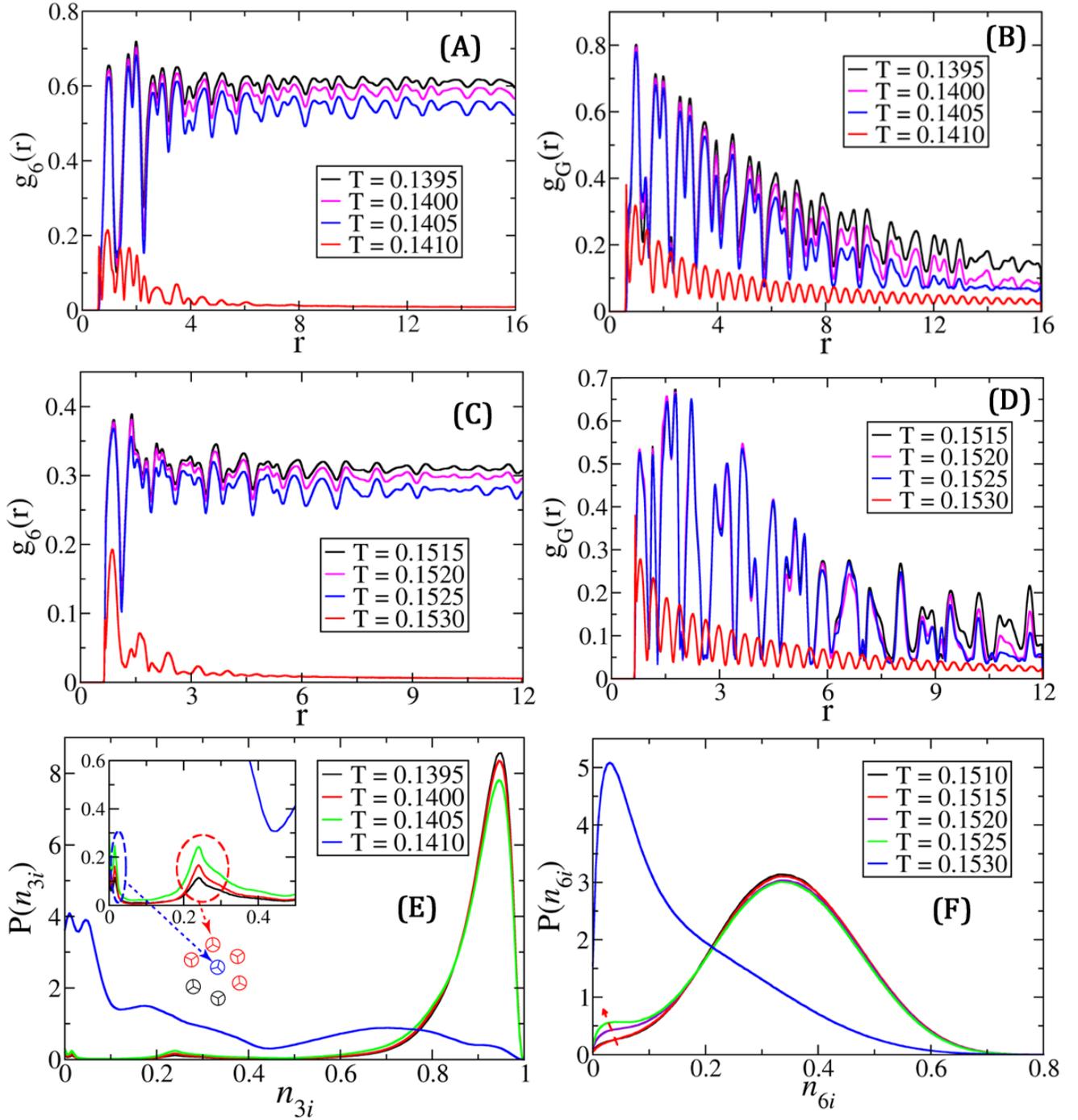

**Figure 5. (A)** Bond orientational correlation function, $g_6(r)$, for different temperatures across the transition at P = 0.19 for $\lambda_{an} = 0.1$. Note that the absence of quasi-long-range bond order (quantified by algebraic decay of $g_6(r)$) and thus appearance of intermediate phase during melting. **(B)** Translational correlation function, $g_G(r)$, for different temperatures at P = 0.19 **(C)**

$g_6(r)$, at different temperatures at P = 0.19 for $\lambda_{an} = 0.4$. We note the presence of weak quasi-long-range bond order (quantified by algebraic decay of $g_6(r)$). **(D)** $g_G(r)$, for different temperatures at P = 0.19 for $\lambda_{an} = 0.4$. Note that the quasi-long-range translational order for solid (quantified by algebraic decay of $g_G(r)$) and short range order for liquid. **(E)** Distribution of local 3-fold bond orientational parameter at different temperatures for $\lambda_{an} = 0.1$ is shown. In inset we have shown the expansion of low order parameter region. **(F)** Distribution of local 6-fold bond orientational parameter at different temperatures for $\lambda_{an} = 0.4$ is shown. Local *m*-fold bond orientational order is defined as $n_{mi} = \left| \psi_{mi}^* \langle \psi_{mj} \rangle_j \right|$, where subscript *j* represents the nearest neighbors of $i^{th}$ particle and $\langle ... \rangle_j$ represents the average over all nearest neighbor particles. Note the emergence of trimodal order parameter distribution for $\lambda_{an} = 0.1$ and bimodal for $\lambda_{an} = 0.4$ near the transition point.

According to KTHNY theory, in the case of melting of triangular solid, the bond orientational order correlation function ($g_6(r)$) will have an algebraic decay ($g_6(r) \sim r^{-\eta_6(T)}, \eta_6 = 18k_BT/\pi K_A$, where $K_A$ is the Franks constant associated with the distortion of bond orientational field) in the intermediate hexatic phase and exponential decay in the liquid phase. In present system, no algebraic decay for $\lambda_i = 0.1$ and weakly algebraic decay for $\lambda_{an} = 0.4$ of bond orientational order correlation function indicates the absence of orientationally anisotropic intermediate phase. One expects to recover KTHNY scenario of melting in the limit of vanishing anisotropic part of interaction energy or specific interactions. We find signatures of emergence of weak algebraic decay of the orientational correlation function (shown in **Figure 5(C)**).

To understand the factors responsible for the decay of $g_6(r)$, we have also plotted the distribution of local bond orientational order at different temperatures (near transition temperature) for both $\lambda_{an} = 0.1$ and $\lambda_{an} = 0.4$ in **Fig. 5(E)** and **Fig. 5(F)**. Unlike in three dimensional systems (3D), in 2D the local symmetry of fluid and crystal phases (near transition temperature) are similar and thus it is not possible to separate unambiguously by just defining the local bond orientational order parameter. In this work we have used a local *m*-fold bond orientational order parameter, first introduced by Grier and coworker [29] and defined as $n_{mi} = \left| \psi_{mi}^* \langle \psi_{mj} \rangle_j \right|$, where *j* presents the nearest neighbors, *m* represents the symmetry of the bond orientation (*m* = 3 for 3-fold symmetry and *m* = 6 for 6-fold symmetry) and $\langle ... \rangle_j$ represents the average over nearest neighbors. This local order parameter represents a projection of $\psi_{mi}$ on the mean local bond orientational field of its nearest neighbors and has information up to 2$^{nd}$ neighbor shell of $i^{th}$ particles.

In **Fig. 5(E)** and **5(F)**, we have shown the local order parameter distributions at different temperatures for $\lambda_{an} = 0.1$ and $\lambda_{an} = 0.4$. As shown in the figure, the order parameter defined by Grier and coworker [29] is also not able to separate the solid and liquid phases unambiguously; however, we have observed that the overlap between the two distributions is significantly reduced compared to $|\psi_{mi}|$ distribution. Near the transition point, we observe the emergence of bimodal character in the distribution of $n_{6i}$ for $\lambda_{an} = 0.4$ (**Fig. 5(F)**). The bimodal character in the distribution of $n_{6i}$ is also observed in freezing transition of atomic systems [30]. The emergence of bimodal character indicates the appearance of liquid-like regions/particles in the bulk metastable parent oblique phase. However, in the case of $\lambda_{an} = 0.1$, near transition temperature, the emergence of three peak nature of distribution indicates a more complex nature of defect configurations. We observe three types of defect configurations – 5-membered rings, 7-memebred rings and highly dynamic triangular defects (where one MB particle (blue colored particle on **Fig. 5(E)**) is trapped inside a honeycomb ring). In the figure the blue colored particle belongs to the blue colored circle and the red colored particles belong to the red colored circle in the local order parameter distribution shown in the inset of **Fig. 5(E)**. Even though the triangular defects are frustrated, highly dynamic they have finite life time as the collective fluctuation in HB network is required to destroy these defects. Despite the fact that these defects are trapped inside six-membered cages, triangular defects do not have six-fold symmetry, as it always stays closer to the one of the MB particles (to be stabilized by LJ interaction) that belongs to the cage. As 5-membered and 7-memeberd rings do not destroy the long range bond orientational order, the triangular defects are only responsible for the breakdown of long range orientational order in the honeycomb solid phase. At transition point, the frustrated triangular defects are stabilized by forming quasi-stable four-membered tetratic defects (see supplementary information **Fig. S1**), which leads to the complete breakdown of bond orientational as well as translational order. For $\lambda_{an} = 0.1$, the complex nature of order parameter distribution in liquid phase also indicates the presence of complex topological defects in liquid.

One can also correlate the decay of bond orientational order correlation, $g_6(r)$, to the free energy surface shown in **Fig. (2)**. In the case of $\lambda_{an} = 0.1$, the curvature of the solid basin in free energy surface as well as the free energy barrier at coexistence is large compared to the case of $\lambda_{an} = 0.4$. The stiffness of the free energy basin suppresses the probability of fluctuations responsible for formation of defects (responsible for breakdown of bond orientational order). Thus we observe a single step transition, without signature of any intermediate algebraic decay. In the case of $\lambda_{an} = 0.4$, the free energy surface is relatively flat which makes formation of defects easier. As a result we find the signature of pre-transition weak algebraic decay. As discussed earlier, the free energy barrier decreases on decreasing the relative strength of anisotropic interaction (increasing $\lambda_{an}$), there will be a crossover from strongly $1^{st}$ order to weakly first order or continuous melting transition on increasing $\lambda_{an}$. In the asymptotic limit (vanishing relative strength of anisotropic interaction) the melting scenario will converge to the usual melting scenario of LJ system.

In order to gain more insight into the structure of the solid and liquid phases near the transition temperature, in **Fig. (6),** we have plotted both the radial distribution functions ($g(r)$) and the nearest neighbor bond length distributions ($P(l_{NN})$) for both $\lambda_{an} = 0.1$ (honeycomb to liquid) and $\lambda_{an} = 0.4$ (oblique to liquid). The splitting of first peak of $g(r)$ for $\lambda_{an} = 0.1$ ( see **Fig. 6(A)**) and development of bimodal character in the nearest neighbor bond length distribution (see **Fig. 6(B)**) indicate the emergence of two length scales after melting and is indeed a consequence of the presence of two length scales in the interaction potential. Liquid consists of two transient, coexisting structures – hexagonal rings and thermally excited defects (tetratic, triangular, 5-membered and 7-membered rings). When thermal energy is sufficient to disrupt the hydrogen bonds, LJ (van der Waals) interaction favors the high density configurations. *At transition temperature, first step towards transition is the formation of quasi-stable four coordinated tetratic structures (in supplementary information see* **Fig. S1***) along with other defects.* The tetratic structures are stabilized by both the H-bond (between diagonally opposite MB particles) and LJ interactions. For $\lambda_{an} = 0.4$, (see **Fig. 6(C) and 6(D)**), the splitting of 1$^{st}$ peak of g(r) and the bimodal character of $P(l_{NN})$ in both the solid and liquid phases indicate the presence of two length scales in both oblique solid as well as in the liquid phases (see supplementary information **Fig. S2)**.

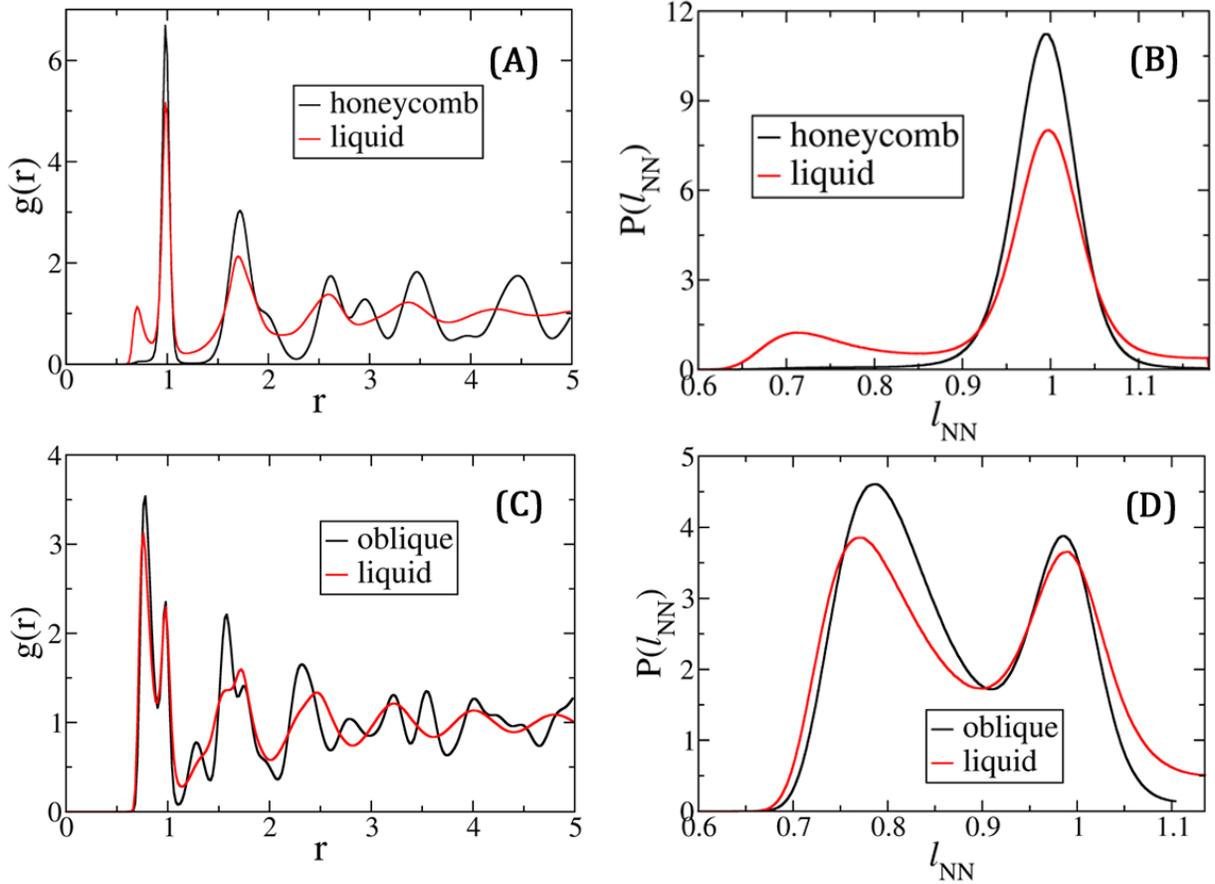

**Figure – 6. (A) Radial distribution functions (g(r)) and (B) nearest neighbor bond length distributions (P(l$_{NN}$)) for honeycomb solid (black line) and liquid (red line) near transition temperature (for $\lambda=0.1$) are shown. We note the splitting of 1$^{st}$ peak of g(r) and bimodal character of nearest neighbor bond length distribution in liquid phase. This indicates presence of two length scales in liquid and arises due to two different length scales of potential. (C) Radial distribution functions (g(r)) and (D) nearest neighbor bond length distributions (P(l$_{NN}$)) in oblique solid (black line) and corresponding liquid (red line) near transition temperature (for $\lambda=0.4$) are shown. Note**

**that the split in 1st peak of radial distribution function and bimodal bond length distribution in both oblique solid and the liquid phases. Plots signify that the liquid has a significant solid-like ordering.**

In this article, we have investigated the role of anisotropic interactions and the strength of isotropic to anisotropic interactions on phase diagram and the nature of melting transition for two dimensional molecular systems. We find that the interplay between the strength of isotropic and anisotropic interactions can give rise to rich phase diagram. The computed solid-liquid phase diagram consists of isotropic liquid and two crystalline phases – honeycomb and oblique. The celebrated KTHNY theory predicts two continuous transitions in the melting of a two dimensional solid. Most of the predictions of KTHNY theory have been verified in simulations of atomistic systems where constituents interact with spherically symmetric potential. The situation can be different in the presence of anisotropic interaction where KTHNY theory appears to be inapplicable. The nature of the transition depends on the relative strength of the anisotropic interaction and a strongly first order melting turns into a weakly first order transition on increasing the strength of the isotropic interaction. This crossover can be attributed to an increase in stiffness of the free energy minima on increasing the strength of the anisotropic interaction. The defects involved in melting transitions of molecular systems are quite different from those known for the atomic systems.

## Methods

The Mercedes-Benz potential has two terms - isotropic Lennard-Jones (LJ) and an anisotropic hydrogen-bond (HB) term, $U(\mathbf{X_i}, \mathbf{X_j}) = U_{LJ}(r_{ij}) + U_{HB}(\mathbf{X_i}, \mathbf{X_j})$, $\mathbf{X_i}$ denotes vectors representing both the coordinates and orientation of $i^{th}$ particle and $r_{ij}$ the distance between the centers of two particles. $U_{LJ}(r_{ij})$ is defined as $U_{LJ}(r_{ij}) = 4\varepsilon_{LJ}\left[\left(\sigma_{LJ}/r_{ij}\right)^{12} - \left(\sigma_{LJ}/r_{ij}\right)^{6}\right]$, where $\varepsilon_{LJ}$ and $\sigma_{LJ}$ are the well-depth and diameter for isotropic LJ interaction and

$$U_{HB}(\mathbf{X_i}, \mathbf{X_j}) = \varepsilon_{HB} G(r_{ij} - \sigma_{HB}) \sum_{k,l=1}^{3} G(\mathbf{i_k}.\mathbf{u_{ij}} - 1) G(\mathbf{j_l}.\mathbf{u_{ij}} + 1), \quad \text{where} \quad G(x) = \exp(-x^2/2\sigma^2).$$

The unit vector $\mathbf{i_k}$ represents the $k^{th}$ arm of the $i^{th}$ particle ($k$ = 1, 2, 3) and $\mathbf{u_{ij}}$ is the unit vector joining the center of molecule $i$ to the center of molecule $j$. The HB parameters are $\varepsilon_{HB} = -1$ and bond length $\sigma_{HB} = 1$. LJ contact distance is $0.7\sigma_{HB}$. Anisotropic parameter is defined as $\lambda_{an} = \varepsilon_{LJ}/\varepsilon_{HB}$. All simulations are performed in constant temperature and pressure ensemble (NPT) at P = 0.10, P = 0.19. The locus of transition points shown in the phase diagram is obtained by first equilibrating the honeycomb lattice consisting 576 particles at T = 0.05 and then gradually increasing temperature in steps of 0.01 for all values of $\lambda_{an}$.

Except phase diagram and free energy surfaces all other simulations are performed with N = 1008. We have first equilibrated the system for $5 \times 10^6$ MC steps and then collected the trajectories for next $5 \times 10^6$ MC steps. For the free energy computation we have used Transition Matrix Monte Carlo (TMMC) [31] with N = 240. All quantities are represented in reduced units such as temperature as $k_B T/\varepsilon_{HB}$ and distance is scaled with hydrogen-bond length, $\sigma_{HB}$.

Translational order parameter ($\psi_T$) is defined as $\frac{1}{N}\left\langle \left\|\sum_{i=1}^{N} e^{i\mathbf{G}.\mathbf{r}}\right\| \right\rangle$, where $\mathbf{G}$ is the first shell reciprocal lattice vector and by definition for prefect crystal $\mathbf{G}.\mathbf{r} = 2\pi$. M-fold bond

orientational order is defined as $\psi_m = \frac{1}{N}\left\langle \left| \sum_{k=1}^{N}\left( \frac{1}{NN}\sum_{j=1}^{NN} e^{im\theta_{kj}} \right) \right| \right\rangle$, where $NN$ is the number of nearest neighbors, $\theta_{kj}$ is the angle of the bond that the $jth$ neighbor makes with the tagged particle and $N$ is the total number of particles in the system. Six-fold bond orientational correlation function is defined as $g_6(r) = \left| \left\langle \psi_6^*(\mathbf{r_i})\psi_6(\mathbf{r_j}) \right\rangle \right|$, where $r = |\mathbf{r_i} - \mathbf{r_j}|$ and $\psi_6(\mathbf{r_k})$ is the six-fold bond orientational order of $k^{th}$ particle. The translational correlation function is given in terms of the Fourier components, $g_G(r) = \left| \left\langle \rho_\mathbf{G}^*(\mathbf{r_i})\rho_\mathbf{G}(\mathbf{r_j}) \right\rangle \right|$, where $r = |\mathbf{r_i} - \mathbf{r_j}|$, $\mathbf{G}$ is the reciprocal lattice vector, $\rho_G = \exp(i\mathbf{G}.\mathbf{r})$ is the Fourier component of the microscopic density $\rho(\mathbf{r})$. The bracket $\langle ... \rangle$ represents particle as well as ensemble average. In this study, bond orientational order susceptibility, $\chi_6$, is defined as $\chi_6 = \left\langle |\psi_6|^2 \right\rangle - \left\langle |\psi_6| \right\rangle^2$, and translation order susceptibility, $\chi_T$, is defined as $\chi_T = \left\langle |\psi_T|^2 \right\rangle - \left\langle |\psi_T| \right\rangle^2$.

**Acknowledgement:** We thank DST, India for partial financial support of this work. B.B. thanks DST for a JC Bose Fellowship.

# Supplementary Information

In **Fig. S1**, we have shown the honeycomb lattice and corresponding liquid phase. Liquid is characterized with two transient coexisting structures – hexagonal clusters and thermally excited defects (tetratic, five membered and seven membered rings). The splitting of $1^{st}$ peak of $g(r)$ and the emergence of bimodal character in nearest neighbor bond length distribution (**Fig. 6A** and **6B**) arises due to thermally excited quasi-stable four-coordinated defects. We note that the tetratic defects are stabilized by both hydrogen bonding with diagonally opposite particles and Lennared-Jones (LJ) interaction.

In **Fig. S2** we have shown the structure of the oblique lattice and the respective liquid phase. Two length scales are clearly evident from the structure of the solid phase. There are string-like arrangements in which intra-string particles are stabilized by vander Waals (LJ) interaction and different string interacting with each other via H-bonding. This interesting structure is a consequence of delicate cooperation between H-bond and vander Waals interaction. The structure of the liquid phase is much more complex than the solid phase. It consists a mixture of vander Waals bonded, hydrogen bonded and non-bonded particles. Bimodal nearest neighbor bond length distribution also suggest the there is two types of local ordering but no global translation of local ordering.

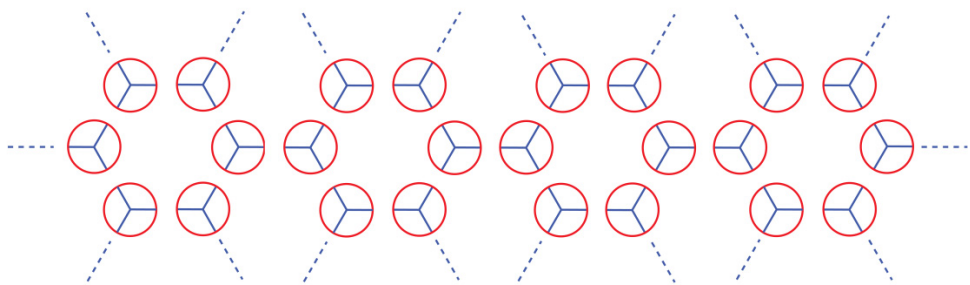

Honeycomb

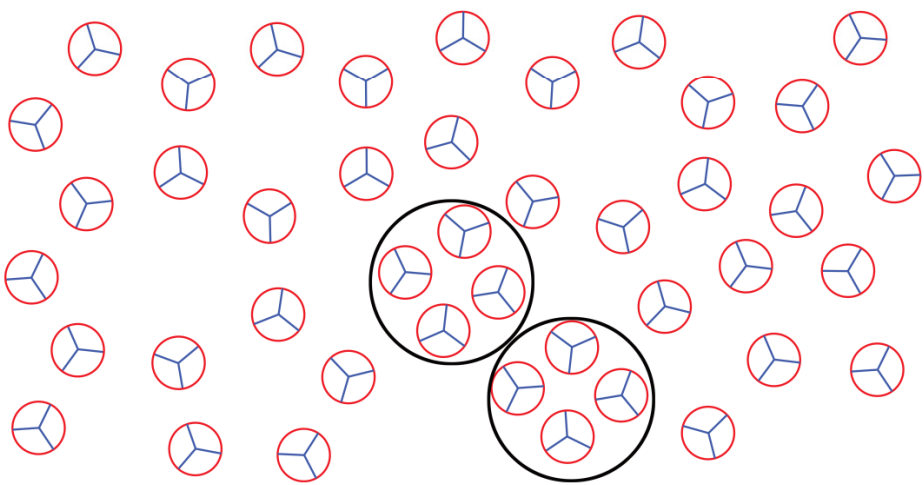

Liquid

**Figure S1. Snapshots for honeycomb solid and corresponding liquid after melting. Liquid has also significant honeycomb like (three coordinated) molecules which are part of five, six and seven membered rings. Black circles indicate the quasi-stable tetratic defects which are stabilized by both H-bond as well as Lennard-Jones interaction. The honeycomb lattice has only one characteristic length scale.**

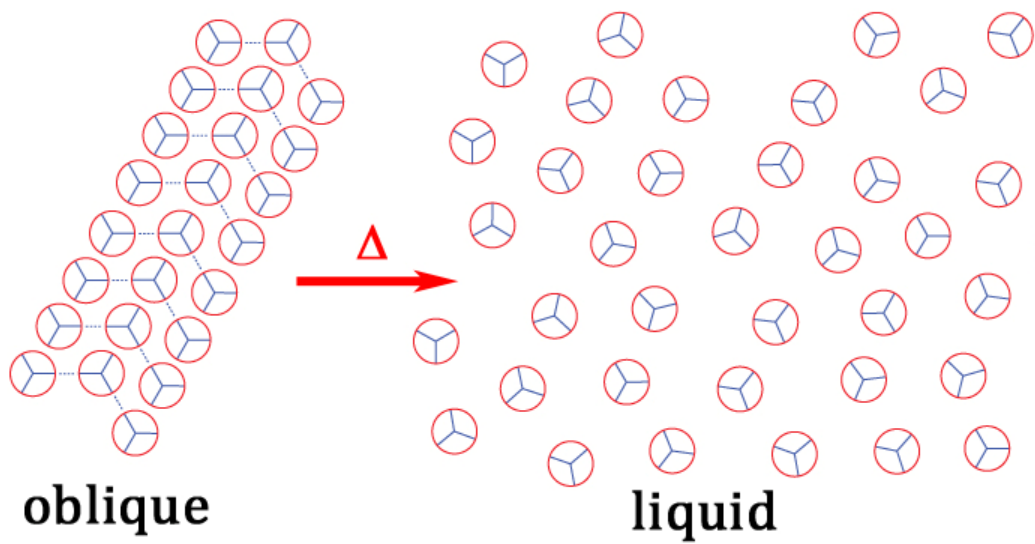

**Figure – S2.** Snapshots of oblique solid (having two length scales) and the liquid configuration after melting are shown. Liquid phase is a mixture of hydrogen bonded, vander Waals (LJ) bonded and non-bonded particles.